\documentstyle[amssymb,aps,preprint]{revtex}

\begin{document}
\title{Vortices in Bose-Einstein condensates confined in a multiply connected
Laguerre-Gaussian optical trap.}
\author{J.\ Tempere and J. T. Devreese}
\address{Departement Natuurkunde, Universiteit Antwerpen UIA, Universiteitsplein 1,
B-2610 Antwerpen, Belgium}
\author{E. R. I. Abraham}
\address{Department of Physics and Astronomy, University of Oklahoma, 440 W. Brooks,
Norman OK 73019, USA}
\date{5/12/2000}
\maketitle

\begin{abstract}
The quantized vortex state is investigated in a Bose-Einstein condensate,
confined in a multiply connected geometry formed by a Laguerre-Gaussian
optical trap. Solving the Gross-Pitaevskii equation variationally, we show
that the criterium for vortex stability is that the interatomic interaction
strength must exceed a critical interaction strength. The time evolution of
a freely expanding Laguerre-Gaussian condensate with a vortex is calculated,
and used to derive the interference pattern of such a condensate overlapping
with a parabolically trapped condensate.
\end{abstract}


\section{Introduction}

Superfluidity, and its characteristic manifestation as a state with
quantized circulation (a vortex), is intimately connected with the
phenomenon of Bose-Einstein condensation, yet the precise relation between
superfluid persistent currents and Bose-Einstein condensation (BEC)\ is only
beginning to be elucidated \cite{Huang}. Before 1995, the link between
superfluidity and BEC was almost exclusively studied in the context of
liquid helium ($^{4}$He and $^{3}$He) \cite{Tilley}, where the study of the
relation between superfluidity and BEC is complicated by the strong
interatomic interactions between the atoms in the liquid. In 1995,
Bose-Einstein condensation was realized in magnetically trapped clouds of
alkali atoms \cite{quattro}. In these novel condensates, the bosonic atoms
are weakly interacting (contrary to the case of liquid helium), and as such
these systems have the potential to shed new light on superfluidity. Soon
after the initial creation of alkali gas condensates, several experimental
groups set out to create a vortex -- a quantum of superfluid circulation --
in this novel system \cite{Varenna98}.

The initial attempts to create a vortex in a condensate by stirring the
trapped condensate with a blue detuned laser beam \cite{Varenna98} were
unsuccessful, even though early theoretical work by Dalfovo {\it et al. }%
indicated that persistent superfluid currents can indeed manifest themselves
in Bose-Einstein condensates as vortices \cite{DalfovoPRA53}. Subsequent
analysis \cite{RoksharPRL79} showed that vortices are unstable in the
simply-connected, not-stirred, spin-polarized condensates formed in the
original experiments \cite{quattro}.

This can be understood as follows. Along the vortex line, the order
parameter of the Bose-Einstein condensate has to vanish. Phrased
metaphorically, a ``hole'' has to be ``drilled'' in the condensate along the
vortex line. In the magnetic trap, the modulus square of the order parameter
of a condensate without a vortex is largest in the center of the trap. As a
consequence, a vortex line through the center of the trap will perturb the
order parameter more than a vortex line at the edge of the condensate
(since, using our metaphor again, it will cost more energy to ``drill''
through the center of the condensate than through its edge). It will be
energetically favorable for the vortex line to be at the edge of the
condensate. In the presence of dissipation, this will cause the vortex line
to migrate to the edge of the condensate so that the vortex condensate will
decay into a non-vortex state. This argument, sketched here with some
roughness, has been worked out with precision by Fetter and co-workers in 
\cite{Fetter}.

Several schemes have been proposed, both theoretically and experimentally,
to stabilize vortices: rotating the trapping potential (analogous to
rotating a bucket containing $^{4}$He) \cite{RoksharPRL79} or stirring the
condensate with an off-resonance laser \cite{JacksonPRL80}, raising the
temperature (to `pin' the vortex in the potential created by the
non-condensate fraction at the center of the vortex) \cite{IsoshimaJPSJ68},
phase imprinting methods \cite{DobrekPRA60}, and various other techniques 
\cite{makevor}. Recently, vortices were created experimentally, both with a
`rotating bucket' experiment \cite{MadisonPRL84} and with the use of a
two-component condensate \cite{MatthewsPRL83}. In the latter experiment, one
of the components of the spinor condensate `pins' the vortex present in the
other component and a Ramsey type interference between the two components is
used to detect the vortex. The long lifetimes of the two-component
condensates in ref. \cite{MatthewsPRL83} is due to similar singlet and
triplet scattering lengths of rubidium, resulting in an anomalously low
inelastic loss rate \cite{MyattPRL78}.

Nevertheless, stable vortices have not yet been realized in a spin-polarized
condensate in non-rotating traps, and new methods of stabilization and
detection must be developed. A promising scheme, based on an analysis
similar to that of Fetter \cite{Fetter}, is the use of multiply-connected
condensates. A candidate trap to create a multiply connected condensates
consists of a red-detuned laser beam in a Laguerre-Gauss mode \cite
{Abraham,KugaPRL78}, which we discuss in section II. If the condensate order
parameter has a toroidal geometry, a vortex line threaded through the
cylindrical symmetry axis of the torus will not perturb the condensate order
parameter strongly, and moreover create a metastability barrier for vortex
decay. One of the goals of this paper is to verify this statement about
vortex stability (in section III). In a condensate with a toroidal order
parameter, the vortex line can no longer be detected as a line along which
the density of Bose-Einstein condensed atoms vanishes. To detect vorticity,
we propose (in section IV) a method based on interference, similar to that
proposed in refs. \cite{TempereSSC108,BoldaPRL81} for simply connected
condensates.

\section{Bose-Einstein condensation in Laguerre Gauss traps}

\subsection{Trapping geometry}

Toroidal confinement for ultracold atoms can be obtained by an optical
dipole trap \cite{StamperKurnPRL80} which consists of a laser beam in a
Laguerre-Gaussian mode. The Laguerre-Gaussian mode $\{n,m\}$ is
characterized by an intensity profile given by \cite{Saleh}:

\begin{equation}
I_{n,l}(r,z)\propto \frac{(2r^{2}/W_{0}^{2})^{l}}{1+(2z/kW_{0}^{2})^{2}}%
L_{n,l}^{2}\left( 2r^{2}/W_{0}^{2}\right) \exp \left\{ -%
{\displaystyle{2r^{2} \over W_{0}^{2}}}%
\right\} ,  \label{LGI}
\end{equation}
where $r$ is the radial distance from the center of the beam, $z$ is the
position along the propagation direction of the beam, $W_{0}$ (referred to
as the `waist parameter') is a parameter controlling the minimal width of
the beam, $k$ is the wave number of the laser, and $L_{n,l}$ is the Laguerre
polynomial of order $\{n,l\}$. Such laser beams have modes which show a node
in the center, and trap the atoms in a cylindrical shell around the axis of
propagation of the beam. The intensity profile of the laser beam in the
Laguerre-Gaussian mode is illustrated in Figure 1. Laguerre-Gaussian (LG)\
laser beams have already been used to successfully trap atoms \cite
{KugaPRL78}, and subsequent theoretical work has shown that toroidal traps
formed by a red-detuned LG\ beam can be loaded from initial conditions
similar to those of conventional magnetic traps \cite{Wright}. Once the trap
has been loaded, one possibility to create a vortex in the trapped gas would
be by a phase imprinting method, which already successfully resulted in the
creation of solitons in condensates \cite{BurgerPRL83}. Alternatively, a
rotating perturbing potential can be used to stir the condensate and set up
persistent flow in the toroidal geometry \cite{JavanainenPRA58}.

In this paper we investigate the properties of the vortex Bose-Einstein
condensate, optically trapped by a laser beam in a Laguerre-Gaussian
propagation mode $\{n,l\}=\{0,1\}$. Along the $z$-axis (the direction of
propagation of the laser beam) an additional magnetic trap \cite{Abraham}
results in a parabolic $z$-axis confinement with frequency $\Omega $, which
prevents the atoms from escaping along the direction of propagation of the
laser beam. This `plugging' of the optical trap was achieved in \cite
{KugaPRL78} by using blue-detuned `plugging beams'. The condensate in the
Laguerre-Gauss geometry \cite{Abraham} will be denoted by `Laguerre-Gaussian
condensate' (LG condensate), in contrast with the condensate in a parabolic
confinement.

The intensity profile of such a Laguerre-Gauss beam in the $zy$-plane (where 
$z$ is the axis of propagation of the laser beam), given by (\ref{LGI}), is
shown in Figure 1. In the remainder of this paper, we use units so that $%
\hbar =m=\Omega =1$ (where $m$ is the mass of the atoms). In these units,
the laser beam parameters for Figure 1 are chosen as follows: $k=2,W_{0}=5$.
Figure 2 illustrates a surface of constant intensity of the laser beam; a
hollow cylindrical shell. This will also be the shape of the cloud of
trapped atoms \cite{Wright}. The extension of the cloud along the $z$-axis
can be tuned by selecting the $\Omega $ frequency of the magnetic trap along
the $z$-axis.

\subsection{Mean-field approach}

Confined Bose-Einstein condensates are well described by a mean-field theory
where the properties of the condensate are derived from a complex function $%
\Psi $. This function (the order parameter) is interpreted as a macroscopic
wave function and obeys the Gross-Pitaevskii equation \cite{pitaevskii}: 
\begin{equation}
-\frac{\hbar ^{2}}{2m}\Delta \Psi +V_{conf}(\rho ,z)\Psi +U_{0}|\Psi
|^{2}\Psi =E\Psi .  \label{GP}
\end{equation}
The term nonlinear in $\Psi $ in (\ref{GP}) arises from the interparticle
interaction potential, which is treated as a contact potential with
scattering length $a_{scat}$ so that $U_{0}=4\pi \hbar ^{2}a_{scat}/m$ where 
$m$ is the mass of an atom. The optical confinement is generated by a
red-detuned Laguerre-Gaussian laser beam in the $\{0,1\}$ mode, with waist
parameter $W_{0}$ and wave number $k$, and propagating along the $z$-axis 
\cite{Saleh}. In addition to this, a harmonic confinement (characterized by
a frequency $\Omega $) is present which confines the atoms along in the $z$%
-direction to a region $z<kW_{0}^{2}$ \cite{Abraham}. The potential energy
corresponding to this geometry is given, in cylindrical coordinates $\{\rho
,\theta ,z\}$, by: 
\begin{equation}
V_{conf}(\rho ,\theta ,z)=-A_{0}\frac{2\rho ^{2}}{W_{0}^{2}}\exp \left\{
-2\rho ^{2}/W_{0}^{2}\right\} +\frac{m\Omega ^{2}}{2}z^{2}.
\end{equation}
In the present treatment, the Gross-Pitaevskii equation (\ref{GP}) will be
solved variationally \cite{variational}. Consider for this purpose the
following two trial functions, expressed in cylindrical coordinates $\{\rho
,\theta ,z\}$: 
\begin{eqnarray}
\psi (\rho ,\theta ,z) &=&{\cal N}_{0}\text{ }\exp \{-\sigma _{0}(\rho -\rho
_{0})^{2}/2-\zeta _{0}z^{2}/2\},  \label{psi} \\
\psi _{\text{v}}(\rho ,\theta ,z) &=&{\cal N}_{1}\text{ }\rho \exp \{-\sigma
_{1}(\rho -\rho _{1})^{2}/2-\zeta _{1}z^{2}/2\}e^{i\theta }.  \label{psiv}
\end{eqnarray}
The variational parameters in the trial functions are $\sigma _{0},\sigma
_{1}$, controlling the width of the functions in the radial direction; $%
\zeta _{0}$,$\zeta _{1}$, controlling the widths in the $z$-direction; and $%
\rho _{0},\rho _{1},$ controlling the radial displacement of the maximum of
the function away from the $z$-axis. ${\cal N}_{0}$ and ${\cal N}_{1}$ are
normalization constants, which are determined by demanding that the norm of
the trial function equals the number of particles in the condensate.

The circulation around a closed loop in the trapped, interacting Bose gas is
defined by $\kappa =\frac{\hbar }{m}\oint \nabla S,$ where $S$ represents
the phase of the order parameter solving the Gross-Pitaevskii equation (\ref
{GP}). The single-valuedness of the order parameter ensures that the
circulation in the trapped Bose gas is quantized: $\kappa =nh/m$ with $%
n=0,1,2,...$. A vortex is present whenever $n>0$. The trial function $\psi
(\rho ,\theta ,z)$ has a constant phase and hence there is no circulation,
no vortex. On the other hand, the phase of $\psi _{\text{v}}(\rho ,\theta
,z) $ changes by $2\pi $ along any closed loop encircling the $z$-axis. Thus 
$\psi _{\text{v}}$ is a trial function for the order parameter of a
condensate with one quantum of vorticity: its circulation equals $\kappa
=h/m $. The trial function $\psi _{\text{v}}$ must have a node along the $z$%
-axis (the factor $\rho $ in expression (\ref{psiv})): if this node were not
present, the variational kinetic energy would diverge along the $z$-axis.

We have used $\psi $ and $\psi _{\text{v}}$ as variational trial functions
to solve the Gross-Pitaevskii equation for a condensate without and with a
vortex, respectively. The resulting variational energy is shown in Figure 3
as a function of $Na_{scat}/a_{HO}$ where $N$ is the number of atoms and $%
a_{HO}=\sqrt{\hbar /(m\Omega )}$. The quantity $Na_{scat}/a_{HO}$ is a
dimensionless measure of the strength of the interaction: upon increasing $%
Na_{scat}/a_{HO}>0$ the interatomic interaction becomes more repulsive. The
energy of the LG condensate with a vortex is larger than the energy of the
LG condensate without a vortex for all interaction strengths investigated.
The difference is smallest for the non-interacting gas and increases
monotonically as $Na_{scat}/a_{HO}$ increases. This means that for all
investigated interactions strengths the vortex LG condensate is either
metastable or unstable. In inset (a) of Figure 3, the optimal value of the
variational parameters $\sigma ,\zeta $ are shown for both trial functions
as a function of the interaction strength. Both the variational trial
function of the LG condensate with vortex and without vortex are broadened
under the influence of the repulsive interactions.

\section{Vortex metastability barriers}

A Laguerre-Gaussian condensate with a vortex can be metastable if there
exists an energy barrier separating this state from a state without a
vortex. In this section we estimate the height of the energy barrier both by
the method proposed by Benakli {\it et al.} \cite{Benakli} for 2D traps, and
by the method proposed by Fetter and co-workers \cite{Fetter}.

\subsection{Hydrodynamic and microscopic instability}

The hydrodynamical and the microscopic instabilities studied by Fetter and
co-workers \cite{Fetter} involve the displacement of the vortex core
relative to the center of the trap. As discussed in the introduction, for a
parabolically trapped condensate it is energetically favorable for the
distance between the vortex core and the center of the trap to increase.
Thus, if dissipation is present, the vortex condensate in the parabolic trap
can decay into a non-vortex condensate through a migration of the vortex to
the edge of the cloud. This effect is difficult to observe experimentally,
since the vortex core can tilt and the image contrast between the vortex
core and the cloud of atoms is reduced \cite{AndersonPRL85}.

To estimate the height of the energy barrier separating the vortex
condensate from the condensate without a vortex in our present case, we
introduce a third variational function to represent the order parameter of a
vortex LG-condensate where the vortex core is at a given distance $R$ from
the cylindrical symmetry axis of the Laguerre-Gauss trapping beam: 
\begin{equation}
\psi _{\text{v}}(R;\rho ,\theta ,z)={\cal N}_{2}\text{ }r(\rho ,\theta )\exp
\{-\sigma _{R}(\rho -\rho _{R})^{2}/2-\zeta _{R}z^{2}/2\}e^{i\Theta (\rho
,\theta )}.
\end{equation}
The function $r(\rho ,\theta )$ gives the distance between the point $\{\rho
,\theta ,0\}$ and the vortex core: 
\begin{equation}
r(\rho ,\theta )=\sqrt{(\rho \cos \theta -R)^{2}+\rho ^{2}\sin ^{2}\theta },
\end{equation}
and the function $\Theta (\rho ,\theta )$ gives the angle between the line
connecting the vortex core with the point $\{\rho ,\theta ,0\}$ and the $x$%
-axis: 
\begin{equation}
\Theta (\rho ,\theta )=\arctan \left( \frac{\rho \sin \theta }{\rho \cos
\theta -R}\right) .
\end{equation}
The parameters $\{\sigma _{R},\rho _{R},\zeta _{R}\}$ are determined
variationally for every studied distance $R$ between the vortex core and the
center of the trapping geometry (the axis of propagation of the LG beam),
and ${\cal N}_{2}$ is a normalization constant. The limit $R\rightarrow 0$
retrieves the results for the vortex condensate studied earlier. In the
limit $R\rightarrow \infty $, the result tends to the result for the
condensate without a vortex. Using the variational approach, we calculate
the energy $E(R)$ of the displaced vortex state for any intermediate $R$. If
the energy $E(R)$ decreases monotonously with increasing $R$, the vortex is
unstable - in the presence of dissipation the vortex condensate will decay.
If there exists a maximum energy $E(R_{\text{max}})$ for an $R_{\text{max}}$
different from zero, there is an energy barrier $E_{\text{barrier}}=E(R_{%
\text{max}})-E(R=0)$ which will hinder the vortex decay.

In Figure 4, the energy barrier $E_{\text{barrier}}$ is shown as a function
of $Na_{\text{scat}}/a_{\text{HO}},$ the dimensionless measure of the
interaction strength used in the previous section. The calculations were
performed for a configuration so that $A_{0}=5$ $\hbar \Omega ,$ $%
W_{0}^{2}=20$ $a_{HO}^{2}$. We found that a metastability barrier exists ($%
E_{\text{barrier}}>0$) for $Na_{scat}/a_{HO}>1.6\pm 0.1.$ This means that
vortices are metastable only if the effective interaction strength
(controlled by the scattering length, the number of atoms and the
confinement strength) is large enough: the metastability is induced by
interactions. Upon further increasing the interaction strength above the
threshold value, the metastability barrier increases. The variational energy 
$E(R)$ is shown as a function of $R$ in the insets: once for an interaction
strength such that the vortex is not stable, and once for a situation where
the vortex is metastable. A reasonable estimate of the experimental
parameters required to realize this trapping geometry gives $a_{HO}=4$ $\mu $%
m and $\Omega =50$ Hz. This implies that for $^{87}$Rb, the critical number
of trapped atoms necessary to make vortices metastable is of the order of $%
10^{3}$. Since one can reasonably expect to trap more than $10^{5}$ atoms in
the trap, vortices created in a Bose gas in the confinement potential
mentioned above, will be well into the metastable regime ($%
Na_{scat}/a_{HO}\approx 625$ for $N=500000$).

\subsection{Uniform transition to a non-vortex state}

The other method which we used to study the metastability of the vortices in
Laguerre-Gauss condensates, was proposed by Benakli {\it et al.} \cite
{Benakli} for 2D traps with a an axial hole, punctured by an off-resonance
laser beam. In this method, a trial solution for the Gross-Pitaevskii
equation is constructed as a superposition between the vortex state and the
state without a vortex : $\Psi (\rho ,\theta ,z)=C_{0}\psi (\rho ,\theta
,z)+C_{1}\psi _{\text{v}}(\rho ,\theta ,z)$ where $%
|C_{0}|^{2}+|C_{1}|^{2}=1. $ Adapting the coefficients $C_{0}$ and $C_{1}$
of this superposition, the function changes from that of a LG condensate
with a vortex to that without a vortex. If the energy of the intermediate
states is found to be higher than the energy of the vortex state, this
constitutes an energy barrier against uniform transition from vortex state
to the state without a vortex, making the former metastable.

Using the variationally optimized trial functions for the order parameter of
a LG condensate with a vortex and without a vortex, we found with the method
of Benakli {\it et al.} that such a metastability barrier exists for $%
Na_{scat}/a_{HO}>1.9\pm 0.1.$ The dependence of the energy barrier on $%
Na_{scat}/a_{HO}$ is shown in Figure 5. In the inset of Figure 5, the
variational energy of the trial function $\Psi $ is shown as a function of $%
|C_{0}|^{2}$ for an interaction strength below (a) and above (b) the
threshold value for metastable vortices. The calculations were performed for
a configuration so that $A_{0}=5$ $\hbar \Omega ,$ $W_{0}^{2}=20$ $a_{HO}$.
Note that the method of ref. \cite{Benakli} followed here does not describe
the decay of vortices mediated by the excitations, for example by nucleation
of vortex loops. However, for a two-dimensional system punctured by an off
resonance laser beam, Benakli {\it et al.} \cite{Benakli} could show that
excitation-mediated decay of vortices only becomes appreciable for vortices
in higher angular momentum states. Furthermore, for a purely one-dimensional
toroidal system, Kagan {\it et al.} \cite{KaganPRA61} showed that for a
weakly interacting system at low temperature, the relaxation time for the
decay the superfluid persistent current due to phonons is strongly
suppressed.

Both the method of Fetter \cite{Fetter} and the method of Benakli {\it et al.%
} \cite{Benakli} give qualitatively the same result, namely that above a
critical strength of interaction, expressed by $Na_{scat}/a_{HO}$, the
vortex state of the LG condensate is stabilized by a metastability barrier.
Quantitative differences remain between the two methods, which is to be
expected since they describe different possible mechanisms of vortex decay.
The general conclusion -- the stabilization of the vortex state due to the
toroidal geometry -- is further supported by a calculation for a vortex
condensate in a mexican hat potential \cite{SalasnichPRA57}.The
metastability barrier calculated as by the method of ref. \cite{Fetter} is
smaller than the barrier calculated used the method of ref. \cite{Benakli}
for $Na_{scat}/a_{HO}\gtrsim 3.5$, which indicates that the latter method 
\cite{Benakli} overestimates the real metastability barrier.

\section{Interference and detection of vorticity}

Phase coherence, a property necessary for the existence of vortices, was
demonstrated experimentally using interference experiments \cite{andrews}.
Two parabolically trapped condensates, displaced by a given distance, are
allowed to expand freely. In the region where the two expanding condensates
overlap, an interference pattern is observed. This type of experiment has
also been proposed to observe vorticity in parabolically trapped condensates 
\cite{TempereSSC108,BoldaPRL81}: the presence of a vortex leads to an
observable {\it edge dislocation }in the pattern of otherwise parallel
interference fringes.

Consider a parabolically trapped condensate positioned on the symmetry axis
of the LG condensate. The LG condensate forms a cylindrical shell
surrounding the parabolically trapped condensate in the center. As the
trapping potentials are switched off, both condensates will expand and
exhibit an interference pattern in the region of overlap. This interference
pattern is the subject of the present section.

To find the function representing the order parameter at a given time $t$
after the start of the free expansion, the original function $\psi _{\text{v}%
}(\rho ,\theta ,z;t=0)$ is expanded in free particle eigenfunctions. These
eigenfunctions acquire a phase factor as time elapses, so that $\psi _{\text{%
v}}$ at a time $t$ is given by the resummed expansion with the `time
evolved' eigenfunctions. More explicitly, for the LG condensate with a
vortex, first the coefficients of the plane wave expansion are evaluated: 
\begin{equation}
\psi _{\text{v}}(\rho ,\theta ,z;t=0)=\int c({\bf k})\frac{e^{i{\bf k}.{\bf r%
}}}{(2\pi )^{3/2}}d{\bf k,}
\end{equation}
\begin{eqnarray}
c(k_{\rho },\phi ,k_{z}) &=&\int \psi _{\text{v}}(\rho ,\theta ,z)\frac{\exp
\{-ik_{\rho }\rho \cos (\theta -\phi )-ik_{z}z\}}{(2\pi )^{3/2}}d{\bf k} \\
&=&\frac{\exp \left\{ -k_{z}^{2}/2\zeta \right\} }{\sqrt{2\pi \zeta }}\times
e^{i(\phi +\pi /2)}%
\displaystyle\int %
\limits_{0}^{\infty }d\rho \text{ }\rho ^{2}J_{1}(k_{\rho }\rho )\exp
\{-\sigma (\rho -\rho _{0})^{2}/2\}.
\end{eqnarray}
In this expression $J_{1}(x)$ is the Bessel function of first order of the
first kind and the wave number ${\bf k}$ is expressed in cylindrical
coordinates $\{k_{\rho },\phi ,k_{z}\}$. As time elapses, the free particles
eigenfunctions acquire a phase factor $\exp \{i\hbar k^{2}t/(2m)\}$. The
function at time $t$ after the start of the free expansion is found by
resumming the eigenfunctions at time $t$ : 
\begin{eqnarray}
\psi _{\text{v}}(\rho ,\theta ,z;t) &=&\int c({\bf k})\frac{e^{i{\bf k}.{\bf %
r}+i\hbar k^{2}t/(2m)}}{(2\pi )^{3/2}}d{\bf k} \\
&=&\sqrt{\frac{1}{1+i\zeta t}}\exp \left\{ -\frac{\zeta z}{2(1+i\zeta t)}%
\right\} \times e^{i\theta }  \nonumber \\
&&\times 
\displaystyle\int %
\limits_{0}^{\infty }dk_{\rho }\text{ }(\rho ^{\prime })^{2}\frac{%
-J_{1}(\rho ^{\prime }\rho /t)}{t}e^{-\sigma (\rho ^{\prime }-\rho
_{0})^{2}/2}\exp \left\{ i\frac{\rho ^{2}+(\rho ^{\prime })^{2}}{2t}\right\}
.
\end{eqnarray}
The time evolution for the free expansion of a condensate prepared in a
parabolic confinement is derived analogously \cite{TempereSSC108}. The total
measured density generated by the two condensates is then given by $|\psi _{%
\text{v}}(\rho ,\theta ,z;t)+\psi (\rho ,\theta ,z;t)|^{2}$ - remember that
each condensate function is normalized to the number of particles in the
given condensate. Several time frames of the resulting evolution of the
density of the expanding condensates are shown in Figure 3. The frames in
Figure 3 show a cross-section of the density along the xy-plane.

Immediately after the traps are switched off, the density is that of a
cylindrical, Laguerre-Gaussian condensate with a parabolically trapped
condensate in the middle. As time goes by, both condensates expand: the
parabolically trapped condensate expands radially, and the cylindrical shell
of the LG condensate broadens. As the expanding condensates start to overlap
the fringe pattern appears. If the LG condensate does not contain a quantum
of superfluid circulation (i.e. no vortex), the interference pattern
consists of a series of concentric circles with linearly increasing radius.
If however the LG condensate does contain a vortex, the interference pattern
is an Archimedean spiral. At higher vorticity, the number of arms in the
Archimedean spiral equals the number of vortex quanta in the LG condensate.

Figure 3 shows that, as a function of time, the spiral interference pattern
rotates around the cylindrical symmetry axis of the trapped condensates,
with a frequency of the order of the frequencies characterizing the
parabolic approximation to the trapping potentials. Furthermore, as can be
seen from Figure 3, the distance between two successive windings of the
spiral increases as time increases. Hence, to detect the spiral interference
pattern, the density has to be measured on a time scale shorter than that
given by the inverse of the characteristic trapping frequencies.

\section{Conclusions}

In conclusion, we have studied the properties of vortices in a condensate in
an optical trap generated by a laser beam in a Laguerre-Gaussian mode and
proposed a method to detect these vortices. The energy of the vortex state
was calculated variationally in a mean-field framework, and a threshold
interaction strength was found beyond which there exists a metastability
barrier stabilizing the LG vortex state against a transition to a non-vortex
state. Both the hydrodynamic instability \cite{Fetter} and the uniform
transition to a non-vortex state \cite{Benakli} were considered in the
investigation of the vortex stability. Below the critical interaction
strength the metastability barrier vanishes, irrespective of the fact that
the present confined geometry allows for the vortex core to be positioned in
a region where the condensate is nonzero.

The time evolution under free expansion of a LG condensate was derived, and
subsequently used to calculate the interference pattern which arises from
the overlap of an expanding LG condensate with an expanding parabolically
trapped condensate in its center. When superfluid circulation is present,
the interference pattern consists of an Archimedean spiral, clearly distinct
from the series of concentric cylindrical fringes which arise in the
interference pattern when no vorticity is present. The observation of such a
spiral interference pattern would constitute clear evidence for vorticity in
a spin-polarized Bose-Einstein condensates.

\section*{Acknowledgments}

We thank F. Brosens and L. F. Lemmens for intensive discussions. Also
discussions with T. Kuga are gratefully acknowledged. This work is performed
within the framework of the FWO\ projects No. 1.5.545.98, G.0287.95,
9.0193.97, G.0071.98 and WO.073.94N (Wetenschappelijke
Onderzoeksgemeenschap, Scientific Research Community of the FWO on ``Low
Dimensional Systems''), the ``Interuniversitaire Attractiepolen -- Belgische
Staat, Diensten van de Eerste Minister -- Wetenschappelijke, Technische en
Culturele Aangelegenheden'', and in the framework of the GOA BOF\ UA 2000
projects of the Universiteit Antwerpen. One of the authors (J.T., aspirant
bij het Fonds voor Wetenschappelijk Onderzoek - Vlaanderen) acknowledges the
FWO for financial support.

\newpage

\section*{Figure captions}

\bigskip

{\bf Figure 1 :} The Laguerre-Gauss (LG) trap is an optical dipole trap
consisting of a red-detuned laser beam. A typical intensity profile of the
laser beam in the LG-mode $\{0,1\}$ is shown in this figure, in a cross
section through the axis of propagation of the beam (the $z$-axis). The
atoms will feel the optical dipole force attracting them to the (toroidal)
region of highest intensity of the laser beam.$\sqrt{20}=\allowbreak
4.\,\allowbreak 472\,14$

\bigskip

{\bf Figure 2} :A typical surface of constant intensity of the laser beam in
the Laguerre-Gauss (LG) mode $\{0,1\}$ is shown. The atoms in the
red-detuned LG beam will be attracted to a toroidal or cylindrical region
such as that within the surface of Figure 2. An additional parabolic
magnetic confinement potential can be added in the $z$-direction \cite
{Abraham,KugaPRL78}.

\bigskip

{\bf Figure 3} : The variational result for the energy of the condensate in
the Laguerre-Gaussian optical trap is shown as a function of the interaction
strength. The full curve shows the energy of the condensate without a
vortex, the dashed curve shows the energy with a vortex. The inset depicts
the results for the variation parameters $\sigma _{0},\sigma _{1}$ in the
trial function for the condensate with vortex (dashed line) and without
(full line), as a function of interaction strength. The trapping (beam)
parameters were chosen as follows: $A_{0}=5$ $\hbar \Omega ,$ $W_{0}^{2}=20$ 
$a_{HO}^{2}$. In all graphs of this Figure, energies are expressed in units $%
\hbar \Omega $ and lengths in units $a_{HO}=\sqrt{\hbar /(m\Omega )}$.

\bigskip

{\bf Figure 4} : The energy barrier for removing a vortex from a condensate
in a Laguerre-Gaussian optical trap, is depicted as a function of
interaction strength. This metastability barrier $E_{\text{barrier}}$ was
calculated using the formalism of Fetter and co-workers \cite{Fetter}. In
this formalism, $E_{\text{barrier}}$ is found by deriving the energy $E(R)$
as a function of the distance $R$ between the vortex core and the center of
the trapping potential, here the axis of propagation of the laser beam. This
is illustrated in the insets. For interaction strengths lower than a
critical value $Na_{scat}/a_{HO}<1.6\pm 0.1$ the barrier vanishes and
vortices in Laguerre-Gaussian condensates are unstable with respect to the
non-vortex state. For interaction strengths above this critical value, a
metastability barrier exists.

\bigskip

{\bf Figure 5} : The energy barrier for a uniform transition from a vortex
condensate to a condensate without a vortex, in a Laguerre-Gaussian optical
trap, is depicted as a function of interaction strength. The energy barrier
for this mechanism was calculated by the method of Benakli {\it et al. }\cite
{Benakli}. In the inset, the energy per particle of the variational function 
$\Psi =C_{0}\psi +C_{1}\psi _{v}$ is given as a function of $|C_{0}|^{2}$,
for two different interaction strengths: one below (a) and one above\ (b)
the threshold for metastability of vortices, and the relation to the energy
barrier for a uniform transition is shown.

\bigskip

{\bf Figure 6}\ : Several time frames in the evolution of freely expanding,
overlapping condensates are shown. The gray scale represents the density in
a cross-section orthogonal to the symmetry axis of the trap (the direction
of propagation of the Laguerre-Gaussian laser beam), with black being the
maximum density and white the minimum density. The initial situation is
depicted in the top left panel: a parabolically trapped condensate in the
center is surrounded by a cylindrical shaped Laguerre-Gaussian condensate
containing a vortex. The time evolution of the density in the region within
the rectangle in the top left panel is shown in more detail in the
subsequent panels. When the confinement of both condensates is switched off,
they expand and overlap - the density in the gap between both condensates
increases - and a spiral interference pattern is formed. Units are chosen so
that lengths are expressed in $a_{HO}=\sqrt{\hbar /(m\Omega )}$ and the time
is in units of $1/\Omega $.

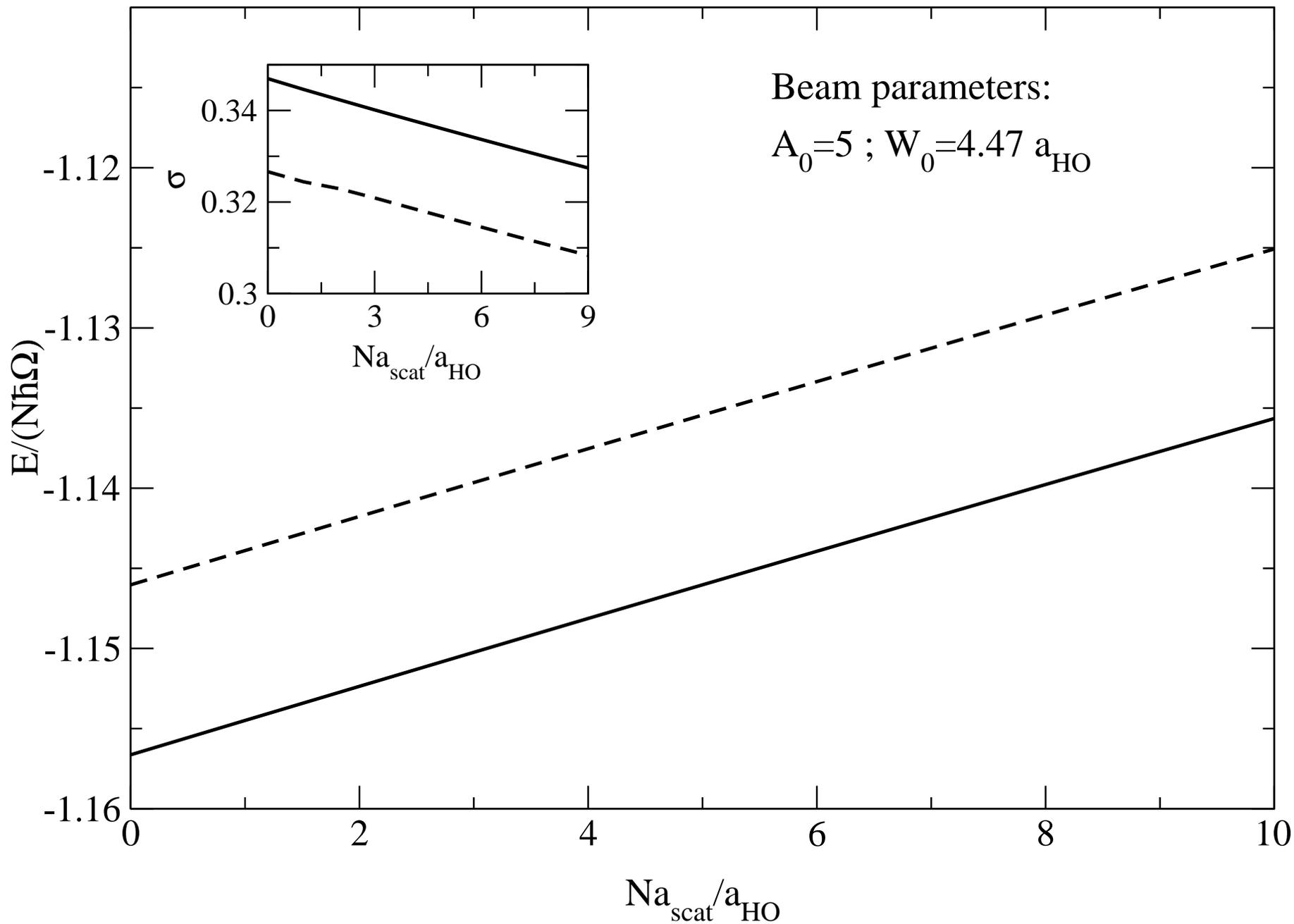
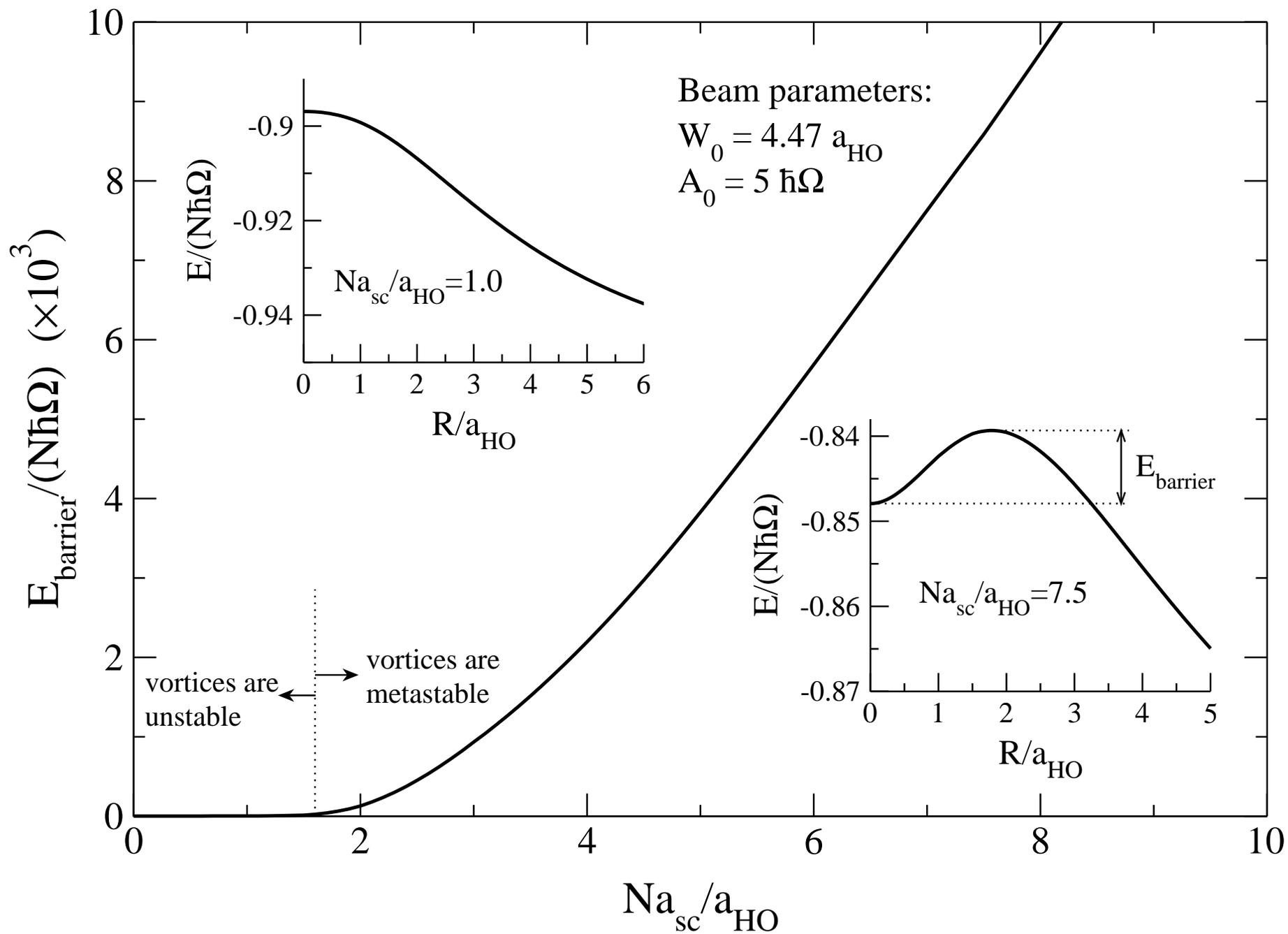
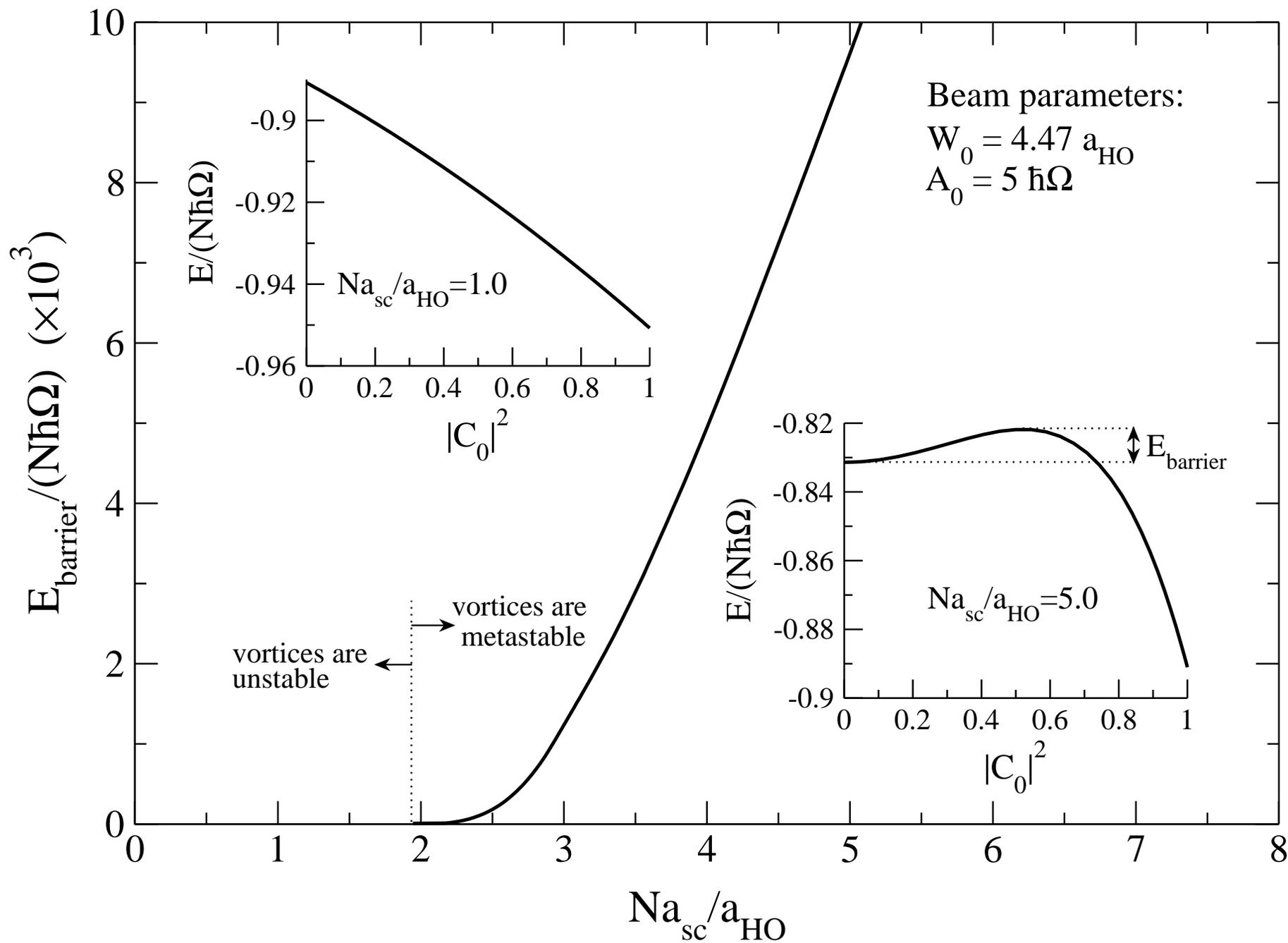
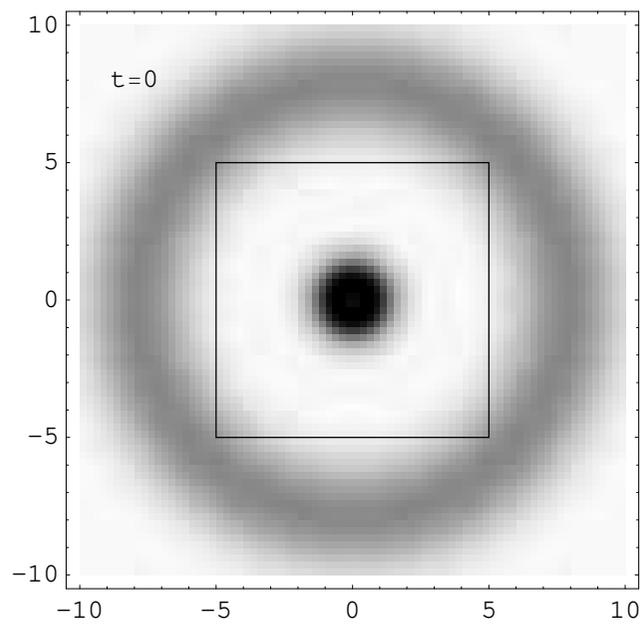
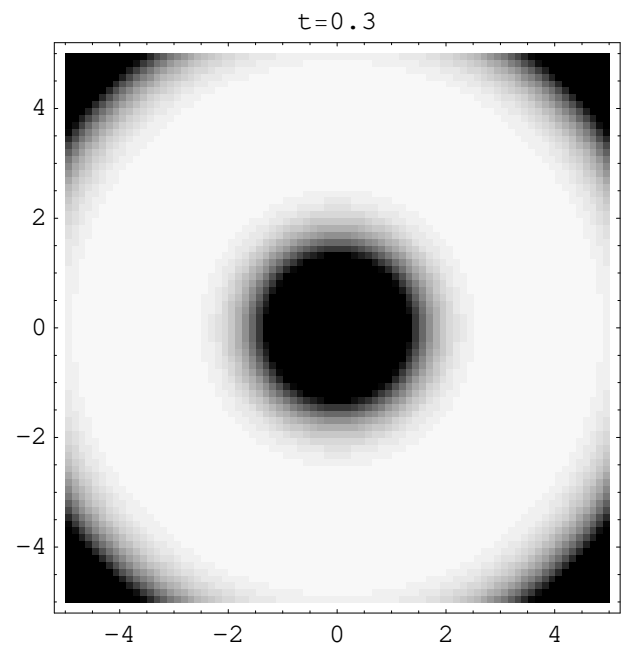
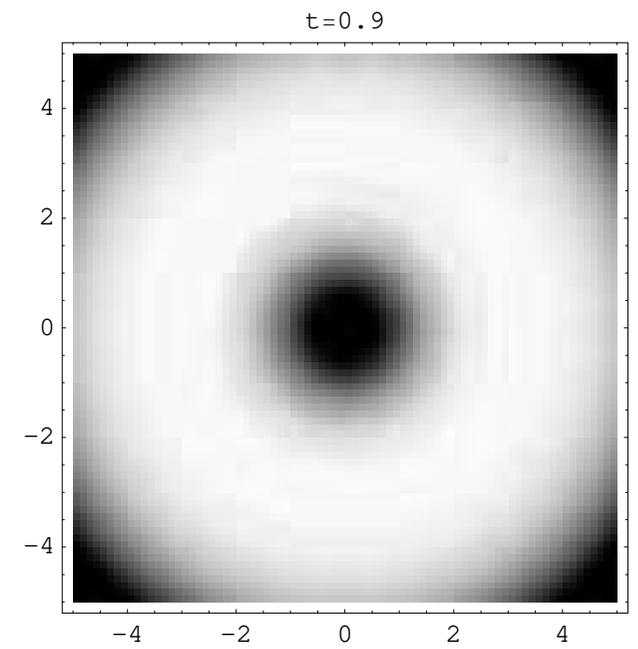
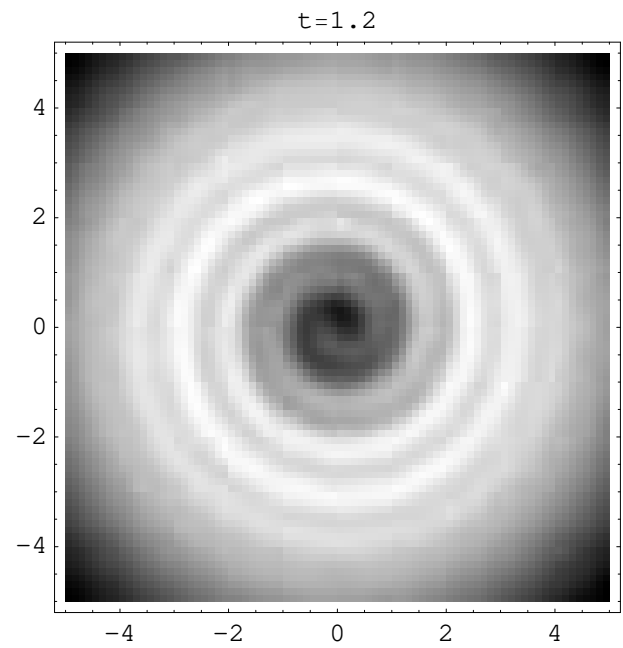
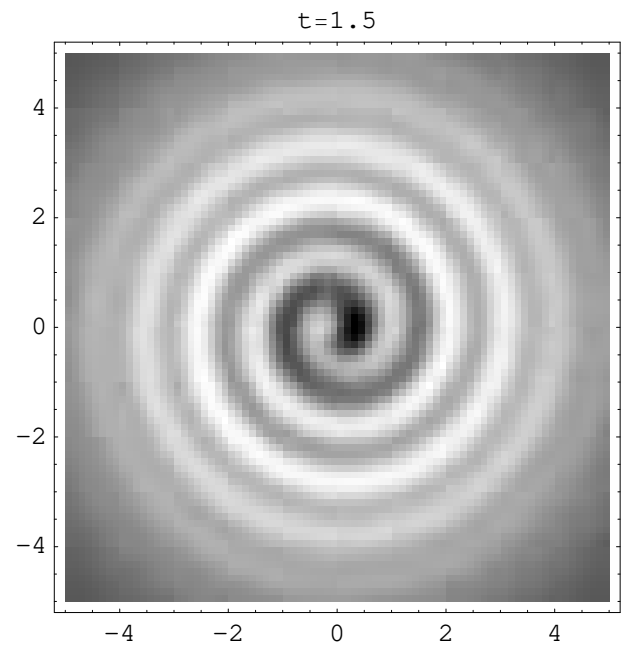
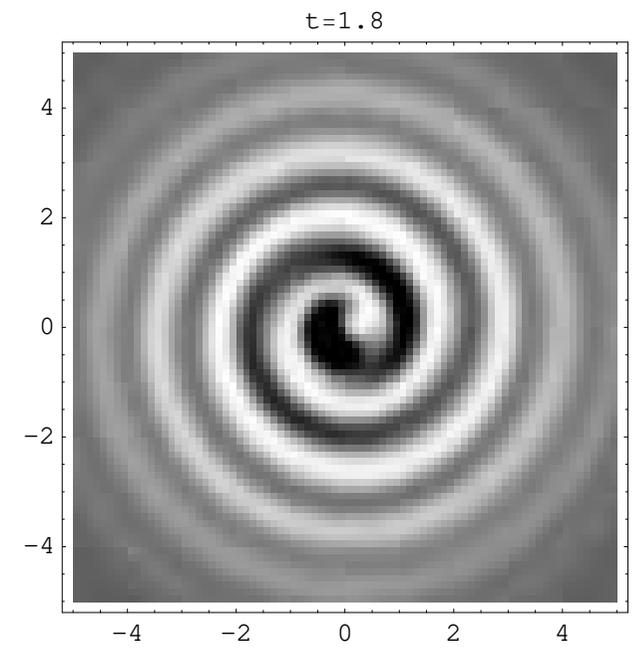
\end{document}